# Numerical simulation of decoherence-control processes in Quantum Computers

**Pedro J. Salas**
*Dpto. Tecnologías Especiales, E.T.S.I. Telecomunicación, U.P.M.,
Ciudad Universitaria, 28040 Madrid, SPAIN*
*psalas@etsit.upm.es*

**Angel L. Sanz**
*Dpto. Física Aplicada, E.T.S.I. Telecomunicación, U.P.M.,
Ciudad Universitaria, 28040 Madrid, SPAIN*
*alsanz@fis.upm.es*

October, 17, 2001

### Abstract

Using a numerical simulation of the evolution of a qubit interacting with the environment we show that quantum error detection and correction can work effectively even when the recovery procedure introduces errors.

## 1. Introduction

Interference and parallelism are the two main properties, which, together with an appropriate quantum algorithm, allow a quantum computer (QC) to have its extraordinary power, in such a way that it will be able to solve some problems intractable with classical computers [1]. But to do that it is necessary to handle and preserve coherent superpositions of quantum states. One of the main obstacles to build a real working QC is the control of the decoherence due to errors coming from interaction with the environment [2]. This loss of coherence results in a corruption of the information stored and a decreasing of the computation power. Some errors may come from mistakes in the initial state preparation, others from imperfections in the implementation of logic gates, and there also will be memory errors appearing in the free evolution of the system.

At first sight, error control looks a very difficult task: quantum errors are analogical (errors in the coefficients of a superposition of states) and not digital as classical errors are. The previous step is to realize that one may digitalize those errors into only three different types: bit-flip errors (characterized by the **$\sigma_X$** Pauli matrix), phase errors (characterized by **$\sigma_Z$**), and bit plus phase errors (characterized by **$\sigma_Y$**). This insight will allow us to quantify classical error correction codes, obtaining from them active methods such as quantum error correction codes.

## 2. Quantum error correcting codes

Calderbank, Shor [3] and Steane[4] have proposed a family of codes achieving a fault-tolerance criteria. The simplest example of encoding protocol is the quantum version of the classical Hamming code C = [7,4,3]. The basis of the quantum code are the words from the classical code, so coherent superpositions can be used to construct a quantum code. The encoding is done with the cosets of the classical code. Steane proposed a [[7,1,3]] quantum code: it encodes a "zero" into a logical zero $|0_L>\equiv \mathbf{C}^\perp$ (dual of C) and encodes a "one" into the logical one $|1_L> \equiv \mathbf{C}^\perp \oplus |1111111>$. For the correction procedure Steane prepares an ancilla state $|a> =(|0_L> + |1_L>)/2^{1/2}$ and with its help extracts the classical-type syndrome that will allow to correct the encoded qubit $|Q> =a|0_L> + b|1_L>$.

The method is based on the necessity of a reliable $|0_L>$ state, as the complete $|a>$ state is obtained from it through a one-time step Hadamard rotation.

But not all ancilla´s error are equally dangerous (referred to network of Figure 1). The worst are bit-flip errors: in the synthesis of $a_X$ ($|0_L>$), as they may transform into phase errors when an Hadamard (H) rotation is applied, infecting thus the qubit $|Q>$ (as can be seen in Figure 1) and in the synthesis of $a_Z$ ($|0_L>$), bit-flip errors can infect the qubit through the CNOT gate .

Phase errors in the $|0_L\rangle$ state are not so destructive, as they only produce an incorrect syndrome. An incorrect syndrome is not so faulty because we repeat it three times and take the most repeated one before correcting the qubit. In this way, the procedure is a fault-tolerant one.

In order to control possible bit-flip errors, Steane [5] proposed a network to have a reliable $|0_L\rangle$ state. It has two pieces: the first one prepares the state and the second one (recovery) checks bit-flip errors and rejects bad states. This recovery network includes an eighth qubit where to accumulate a parity qubit employed to reject those ancillas having a bit-flip error. This network can't detect the occurrence of two bit-flip errors in the ancilla synthesis, and they would contaminate the qubit $|Q\rangle$.

The full qubit correction procedure is done in two steps, sketched in Figure 1:

1).- phase errors are corrected using an ancilla state $a_Z(|0_L\rangle)$ (the line in the middle of Figure 1).

2).- bit-flip errors are corrected with the ancilla $|a\rangle$ obtained from a Hadamard rotation applied to $a_x(|0_L\rangle)$ (bottom line in Figure 1).

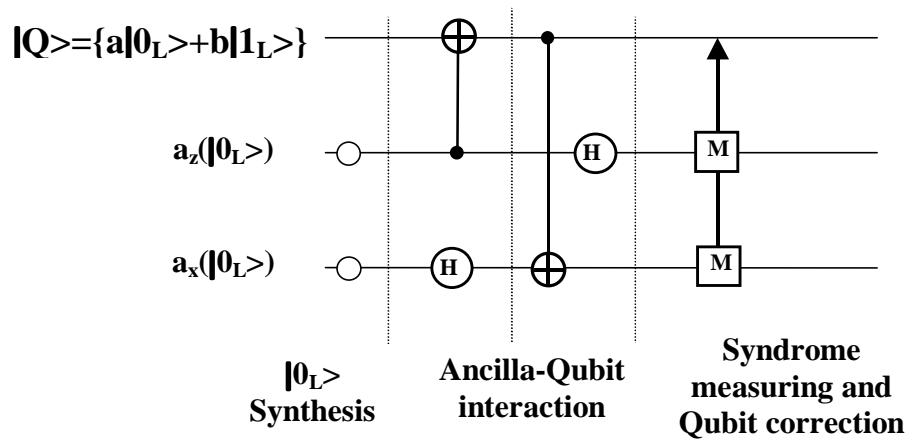

**Fig. 1**. Qubit error correction using a two-step ancilla recovery method

Difficulties during error correction will increase if the recovery procedure introduces itself some errors.

## 3. Numerical simulation

We have done the simulation to correct errors using Steane's quantum code and introducing errors in the recovery method. The isotropic depolarizing channel model is used. Every one of the three quantum errors appears with a probability $\varepsilon/3$ whereas two qubit errors will appear with a probability $\gamma/15$ (note that $I \otimes I$ is not a real error) where $\gamma$ is the error probability in the two-bit gates and $\varepsilon$ is the error probability in a one-qubit gate or memory time step.

In this work our aim is to study the conditions in which a quantum noisy error correction method can control qubit errors. The initial ancilla state $|0_L\rangle$ is considered. The ancilla's quality has been tested by calculating the fidelity as $F = |\langle a|a_{err}\rangle|^2$ (overlapping between the correct ancilla vector $|a\rangle$ and the vector synthesized through a network introducing errors $|a_{err}\rangle$). The results of this calculation appear in Figure 2.a.

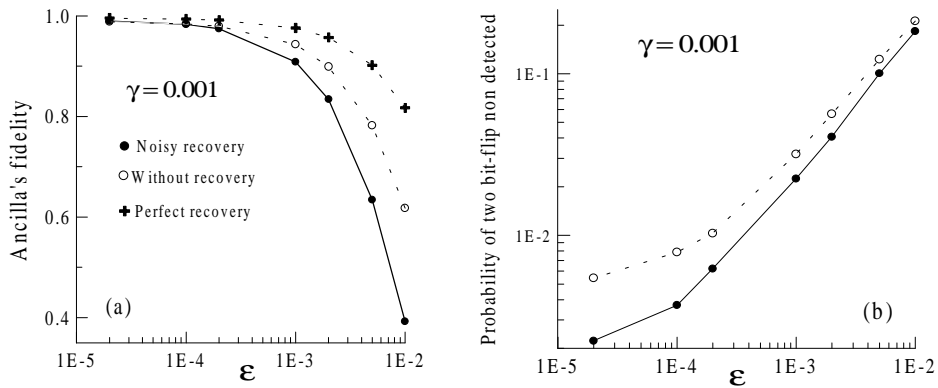

**Figure 2**

Taking only into account the ancilla's fidelity, it seems that no recovery is a better method than a noisy correction method with encoding + correction. This unexpected behaviour originates in the number of noisy gates employed in the recovery ancilla piece.

In fact it is a wrong conclusion, as fidelity is not the best measurement of ancilla´s quality. The errors that may corrupt the qubit we want to correct are two bit-flip non-detected errors, so we should look at their probability $P(\varepsilon,\gamma)$. Figure 2.b plots P ($\varepsilon = 10^{-3}$, $\gamma$) showing that noisy Steane´s error correction actually provides the best ancilla state.

In order to verify if a detection and correction method is reliable, we study the transmission of a ´naked´ qubit $|Q_{naked}\rangle = (|0\rangle+|1\rangle)/2^{1/2}$ propagating along **t** time steps through a quantum channel with evolution errors (of probability $\varepsilon$). Two different methods are used to send the qubit along the channel. First, a naked qubit with no encoding and no error correction. In this case, for small $\varepsilon$ values, fidelity may be estimated by $(1-2\varepsilon/3)^t$. Second, an encoded qubit $|Q\rangle = (|0_L\rangle+|1_L\rangle)/2^{1/2}$ previously prepared without errors. Encoding will allow detecting and correcting errors along time.

In this simulation, correction takes place with the maximum frequency, that is, with only one time step between two subsequent corrections. The results appear in Figure 3 for $\varepsilon = 0.0002$ and different $\gamma$ values. For $\gamma$ small enough, there is a critical time beyond which the application of an encoding + error correction method works effectively to control decoherence.

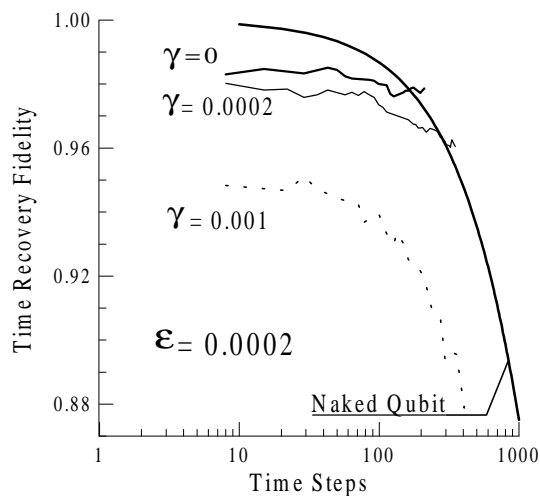

**Fig.3** Time recovery fidelity for different error probabilities.

## 4. Conclusions

Numerical simulation shows that there exists a critical time for certain γ values (corresponding to the crossing points between γ = 0 , γ = 0.0002 and the naked qubit curve) beyond which the noisy recovery procedure is more reliable than the transmission of a naked qubit.

## 5. Acknowledgements

We wish to thank Dr. A. Steane for helpful discussions concerning the model. This work has been supported by Spanish Ministry of Science and Technology Project N. BFM2000-0013.